\documentclass{article}
\usepackage[utf8]{inputenc}
\usepackage[T1]{fontenc}
\usepackage{amsmath,amssymb}
\usepackage{graphicx}
\usepackage{hyperref}
\usepackage{geometry}
\usepackage{caption}
\usepackage{titlecaps}
\usepackage[english]{babel}
\usepackage{csquotes}
\usepackage{listings}
\usepackage{xcolor}
\usepackage{xparse}
\usepackage{pythonhighlight}

\usepackage{float}
\floatstyle{plain}
\newfloat{CodeBlock}{htbp}{loc}[section]

\usepackage{caption}
\title{Fetch.ai: An Architecture for Modern Multi-Agent Systems}
\author{Michael Wooldridge$^1$, Attila Bagoly$^2$, Jonathan J. Ward$^2$, \\ Emanuele La Malfa$^1$, Gabriel P. Licks$^2$}
\date{%
    \small
    $^1$University of Oxford\\%
    $^2$Fetch.ai\\[2ex]%
}

\begin{document}

\maketitle
% \tableofcontents

\section{Introduction}\label{introduction}

While the concept of \emph{intelligent agents} has been around for at least three decades, it has recently come to the fore again, driven in part by the emergence of Large Language Models (LLMs), and the possibility of using these powerful general-purpose AI tools as the ``brains'' for AI agents that act on our behalf. The past few years in particular have witnessed an explosion of interest in such agents. However, this new wave of intelligent agents is largely uninformed by decades of previous experience -- good and bad. In this report, we describe the origins of the agent paradigm, and the main ideas that have informed its development. We go on to look at the recent emergence of agents in the era of LLMs, and describe some recent examples of the approach. We then describe the Fetch.ai agent platform: a state-of-the-art platform for multi-agent operation that is designed from the bottom up to provide a safe, robust, industrial-strength framework for the rapid development and deployment of multi-agent solutions, which leverages both the rich history of multi-agent development, as well as more recent advances in LLM agents.

\section{Agents: The Origin Story}\label{agents-the-origin-story}

The twin concepts of \emph{agents} and \emph{multi-agent system} have
been central to the field of Artificial Intelligence (AI) for more than
three decades. The concepts of agents emerged from several disparate
threads of AI research that came together in the late 1980s and early
1990s.

One key driver was the idea of AI systems interacting with one another.
The history of computing has been marked by a steady trend away from
monolithic, isolated systems to networked, distributed systems. By the
mid 1980s the internet was becoming widely known and used in academic
circles, and the idea of ``distributed AI'' seemed natural. Attention
turned to how networks of communicating AI systems might be designed to
collectively solve problems, for example by sharing tasks or information
\cite{Smith1980}.

Another thread that led to the emergence of the agent paradigm was the
idea of changing the way that we interact with software. Most software
applications like Microsoft Word or Google Chrome are the passive
recipients of our instructions. By and large, they do something only
when we tell them to do something -- when we click on a button on the
app, or select an item from a menu. There is only one agent in this
process: you, the user. Everything happens because you make it happen.
By the late 1980s, researchers were considering the next iteration of
the classic Window-Icon-Mouse-Pointer model for user interaction, which
had been made popular with the release of the Apple Macintosh computer
in 1984. Ideas began to coalesce around the possibility of changing the
way we interact with computers, from one where we essentially tell the
computer what to do, to one where the computer \emph{actively cooperates
with us} on our task, in the same way that a human assistant would. The
idea was thus that the program would act on our behalf -- as our
\emph{agent}. This version of agency led to a swathe of prototype agents
in the late 1990s, assisting users with specific tasks like reading
email and arranging schedules \cite{Maes1994}.

These versions of agency coalesced in the early 1990s. By the mid 1990s,
the field of agents was firmly established, drawing researchers from AI
and a wide range of other disciplines. A ``standard model'' began to
emerge, whereby the field of AI itself was actually defined as being
concerned with constructing software/hardware agents to which we
delegate our preferences, and which then act rationally and autonomously
in some environment in pursuit of those preferences \cite{Russell2020}. The environment may be a virtual one (e.g., a computer game), a
software environment (e.g., a computer operating system or network), or
a physical environment (e.g., a warehouse or a soccer pitch).

\begin{figure}
    \centering
    % \includegraphics[width=0.7\linewidth]{figures/standard-model.jpg}
    % \vspace{0.2cm}
    % \hrule
    % \vspace{0.3cm}
    \includegraphics[width=0.9\linewidth]{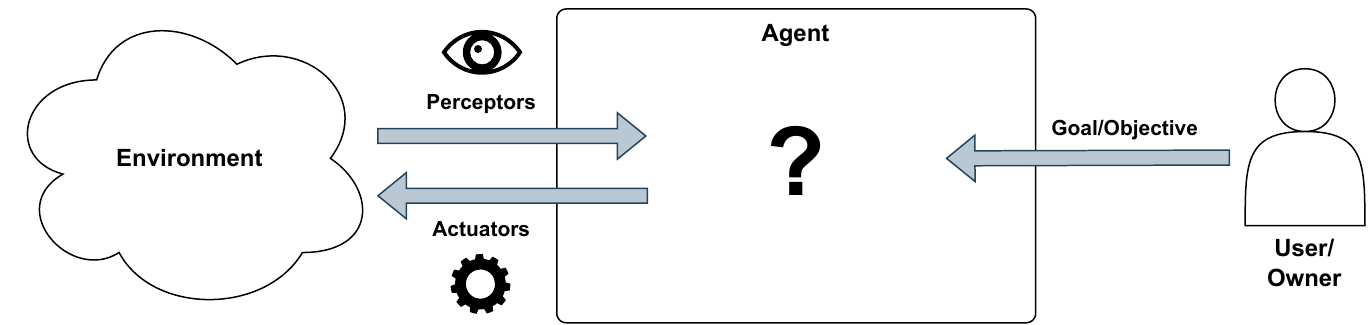}
    \hspace{0.3cm}
    \caption{The standard model. We build an agent to operate on our behalf in some environment---autonomously pursuing our delegated goals/objectives.}
    \label{fig-standardmodel}
\end{figure}

The key challenge for AI, according to the Standard Model, is to
construct a suitable \emph{agent control system}. The classic AI
approach to this problem proposes a \emph{vertical decomposition} of
intelligence into its functional components: perception, learning,
planning/problem solving, reasoning, and so on; typically, such
approaches assume a ``world model'' (the \emph{beliefs} of the agent),
acting as a shared data structure (see Figure~\ref{fig-standardmodel}). We
build an agent to operate on our behalf in some environment -
autonomously pursuing our delegated goals/objectives. While such a
vertical decomposition is natural, it was subsequently criticised for
not giving any indication of how the various components would be
assembled into an \emph{integrated agent architecture}. The field of
\emph{agent architectures} thus emerged, aimed at developing
methodologies for the design of agents in specific domains.

\subsection{Multi-Agent Systems}\label{multi-agent-systems}

If our agent needs to interact with other agents, either human or AI,
then our agent will require \emph{social skills.} As a baseline, it will
need the ability to communicate and to understand what other agents are
saying -- and for interactions with humans, this means some ability to
understand natural languages like English. For agent-to-agent
interaction, however, the early focus was on artificial languages for
agents such as FIPA \cite{FIPA}.

Early attempts to define artificial agent communication languages took
their inspiration from theories of human communication. They assumed
that agents are truly autonomous in the sense that one agent cannot
invoke a method on another, as in object-oriented programming. Instead,
agent communication assumed that agents would communicate via messages
in which they \emph{inform} one another of states of affairs in order to
communicate information, and \emph{request} one another to carry out
some action \cite{JADE}.

A core problem within such languages is that of \emph{shared meaning}:
ensuring that the terms used by one agent are the same as the terms used
by another. To this end, a major enterprise sprang up formally defining
\emph{ontologies} for agents \cite{OWL}. Such
ontologies provide a shared, structured vocabulary that enables agents
to understand and interpret messages consistently. They define concepts,
relationships, and rules within a specific domain, ensuring semantic
interoperability between heterogeneous systems. For example, in
multi-agent systems for e-commerce, an ontology might define terms like
\emph{Product}, \emph{Price}, and \emph{Customer} to facilitate
transactions between agents from different platforms. Similarly, in
healthcare, an ontology such as SNOMED CT ensures that medical agents
accurately exchange patient data by using standardized terms for
diagnoses and treatments. By enforcing a common semantic framework,
formal ontologies reduce ambiguity and enable more effective
coordination, negotiation, and knowledge sharing among autonomous agents
\cite{SNOMED}. Large scale ontologies such as Schema.Org
are now routinely used across the World-Wide Web \cite{Schema}:
Schema.org provides a structured vocabulary for annotating web content
with metadata, enabling search engines and other applications to better
understand and categorize information. For example, it allows a precise
classification of data items on the web, making it possible to
explicitly label data items relating to entities such as names,
addresses, products, events, organizations, and documents.

\begin{figure}[t]
    \centering
    % \includegraphics[width=0.37\linewidth]{figures/agent-architecture.jpg}
    % \hspace{0.3cm}
    % \vrule
    % \hspace{0.5cm}
    \includegraphics[width=0.7\linewidth]{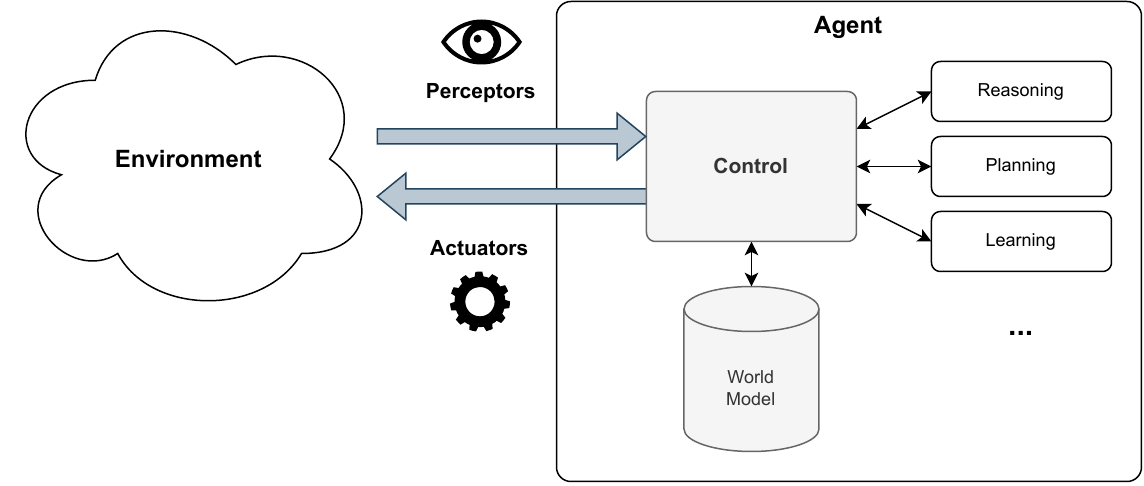}
    \caption{A classic agent architecture, involving a ``vertical decomposition'' of intelligence into its component pieces: reasoning, planning, learning, etc.}
    \label{fig-classicagent}
\end{figure}

While agent communication languages and ontologies provide a substrate
through which agents can communicate, and ontologies provide a framework for sharing meaning, they do not fundamentally address the problem of social skills---and so a huge endeavour began, investigating the problem of equipping agents with these capabilities.
Coordination is the problem of managing interdependencies between agent activities. Dependencies can be positive or negative. A positive dependency occurs when I can carry out an action that in some way supports you. For example, it may be that my activity (e.g., ``clean the house'') completely subsumes yours (``clean the kitchen''): me carrying out my activity relieves you of the need to carry out yours. Alternatively, my activity may subsume part of your activity, in which case I am relieving you of some of your burden. Negative dependencies may also arise. For example, me carrying out my activity may ``clobber'' yours---completely prevent it from taking place. An example might be when I consume a non-renewable resource that you had planned to use as part of your activity. My activity can also impede yours---for example using a printer that you were intending to use, forcing you to wait. Understanding and managing these dependencies became a key activity in multi-agent systems, enabling agents to reason about their activities and how they affect others \cite{LithgowSmith2011}.

Mechanisms for reaching agreement---bargaining and negotiation---were also widely studied. A classic scenario involved agents determining how to share the benefits of cooperation. For example, consider two agents who must each complete a set of tasks. If there are duplications between their tasks, then the agents may be able to mutually benefit by sharing the duplicates amongst themselves to avoid duplication (and hence unnecessary effort). The problem is determining how they should divide the tasks: if both agents want to minimise effort, then from the point of view of both of them, the best thing would be for the other to carry out all the duplicate tasks. Of course this would be the worst outcome from the point of view of the agent doing all the duplicate tasks. Protocols were developed for situations like this that enabled agents to reach agreements through a series of negotiation steps, in which at every step, each agent makes a proposal regarding the division of duplicate labour \cite{RosenscheinZlotkin1994}. A general problem in AI settings like this is that we want to avoid situations in which one agent can gain an advantage on the other through some means. For example, we would like to prevent cases where an agent with more RAM or a more powerful processor available would be able to benefit from this in negotiation: we want the outcome to be guaranteed to be fair to all participants. A range of protocols were developed for doing this, providing mathematical fairness guarantees on outcomes \cite{Fatima2014}. Interestingly, one aspect of these protocols was that it would be possible for one agent to share its program code, and know---with mathematical certainty---that the other agent would not be able to benefit from this. On the contrary, agents would have an incentive to share their program code, to avoid misunderstandings by other agents.

This shared meaning, cooperation, coordination, negotiation, and so on
\cite{Wooldridge2009} and that agent interactions relating to cooperation,
coordination, and negotiation are governed by specified protocols
\cite{Bauer}. To define the semantics of these
protocols, the mathematical apparatus of game theory was widely adopted,
for example with the goal of proving that a negotiation protocol would
achieve outcomes that are ``fair'' to participants under the assumption
that these participants act rationally (for some suitable mathematical
formulation of fairness) \cite{Fatima2014}.

Although there was an immense flowering of interest in multi-agent
systems at the turn of the century, the Standard Model was not widely
adopted. Components of the model were successful in niche applications
(e.g., formal ontologies were instrumental in the emergence of knowledge
graphs \cite{Ontology}), but the idea that multi-agent
systems would be a software paradigm akin to object-oriented programming
did not gain traction.

We can identify several reasons for this.

\begin{itemize}
\item
  \textbf{The preference elicitation bottleneck}. If an agent is going
  to do something for us, then we need to communicate our preferences to
  that agent. The problem is, this is not necessarily easy to do. When
  we communicate with a human assistant our preferences for a travel
  booking (for example), that assistant will make use of rich background
  knowledge of human interaction in order to understand and predict our
  preferences. For example, they would (hopefully) realise that on a
  trip from London to Delhi, a nine-hour stopover in Moscow in order to
  save £10 would not be appropriate or cost effective. The problem is,
  until very recently, we did not have AI that was capable of
  understanding such nuances; and it would be impossible in practice to
  communicate all such nuances explicitly. This difficulty in formally capturing and communicating complex user preferences remains one of the most formidable obstacles to the widespread deployment of computer-aided decision support systems \cite{Braziunas2006}. As a result, users became
  frustrated with a lot of early AI agents - an (in)famous example was
  Microsoft's ``Clippy'' assistant for their Office suite, which
  attracted ridicule in the late 1990s, and was quietly dropped by
  Microsoft from subsequent releases of the flagship Office suite \cite{clippy2025}.
\item
  \textbf{The need for language}. Relatedly, efforts to build practical
  agents were hobbled by the poverty of interfaces. In many settings,
  (e.g., while driving a car), verbal communication is the only
  realistic way for a user to communicate with an agent - but, again,
  until very recently, NLP was simply not sufficiently capable to be a
  realistic option. Much early criticism of Apple's SIRI agent, for
  example, was directed to the fact that it was only capable of
  conversations ``on rails'', on a very narrow set of topics using
  essentially a pre-programmed vocabulary. Users were immediately
  frustrated and disappointed - not least because expectations were set
  high by advertising that highlighted ``AI'' interfaces.
\item
  \textbf{Lack of power}. The lack of power in PCs and handheld devices
  greatly restricted the use of AI techniques in early AI agents. Put
  simply, it wasn't possible to deploy AI powerful enough to build
  practically useful agents until recently.
\item
  \textbf{Not integrating with the software developer toolchain}.
  Perhaps the most important lesson learned from the first wave of AI
  agents was the need to provide agent platforms that seamlessly
  integrate with developer toolchains and workflows. Too many agent
  platforms were clearly academic tools, not addressing the needs or
  working practices of software developers. As a result, developers
  simply ignored them, and looked for other approaches to solve their
  coding issues.
\end{itemize}

\section{The Emergence of LLMs\ldots and LLM-based Agents}\label{sec:emergence-llms-and-agents}

\subsection{The Emergence of LLMs\ldots{}}\label{the-emergence-of-llms}

The development that has catalysed the recent interest in agents has
come from the development of LLMs. These are
based on a technology known as Deep Learning, which uses neural networks
with very large numbers of layers hence ``Deep''. This architecture has
led to enormous progress across all fields of Machine Learning,
initially in machine vision but later in producing super-human
performance in areas such as the game of Go and computer games. The
second important enabling feature of LLMs was the invention and
refinement of the \emph{transformer} architecture, which led to a
step-change in the performance of systems for translation, speech
recognition and generative art.

While all of these breakthroughs provide extremely powerful tools for
the development of agent-based systems, the most important breakthrough
was the emergence of transformer-based LLMs, and
the flagship system in this area was OpenAI's GPT-3 \cite{brown2020}.
GPT-3 represented an unprecedented step-change in capability over its
predecessor GPT-2. The ability of GPT-3 to process and generate
high-quality natural language text immediately attracted attention -- a
range of problems in Natural Language Processing (NLP) that had been
studied for decades were overnight made irrelevant. This in itself was a
remarkable achievement, but for AI researchers, just as important were
the \emph{emergent} capabilities that GPT-3 appeared to have acquired. A
classic example is \emph{few-shot learning}: GPT-3 was able to learn how
to carry out (language-based) tasks based on a very small number of
examples, without specialised training (Brown et al. 2020). But more
than this, early adopters excitedly reported success when applying GPT-3
to a range of classic AI problems such as logical reasoning, commonsense
reasoning, planning, and problems in the theory of mind. If true, this
would suggest it might be possible to leverage LLMs like GPT-3 to solve
a host of previously intractable AI problems. Since then, enormous
efforts have been directed to understanding the extent of LLM
capabilities. Our current understanding is very incomplete and
uncertain, though initial speculation that LLMs were capable of
(zero-shot) abstract reasoning and planning has been debunked
\cite{Kambhampati2024}. Attempts to overcome these limitations through the
use of ``Chain-of-Thought'' prompting, for example, remain tentative;
and fine-tuning approaches have been criticised for essentially
pre-compiling solutions that an LLM then simply recalls. Nevertheless,
LLMs are powerful general purpose AI tools -- and agents are a natural
medium for deployment.

\subsection{\ldots and LLM-based Agents}\label{and-llm-based-agents}

Over the past 18 months, a number of LLM-based agent frameworks have
been proposed. Most of these are essentially simple automation packages:
scripts which make use of LLMs to automate some language-based tasks
(e.g., sending customised messages to all LinkedIn contacts). Work in
this area is largely uninformed by existing multi-agent systems
research, and the extensive experience gained on (e.g.) interaction
protocols.

LLMs are extremely good at understanding the nuances of human preferences through language: and this capability goes a long way to addressing the preference elicitation bottleneck. For the first time, we can build agents that we can communicate with using the full expressive power of natural languages; LLMs can even understand colloquial language and slang. This opens the door to a vast range of applications that will enable us to deploy the power of LLMs in a range of settings.

In addition, though, it is the ability of LLMs to reason and solve problems that has attracted attention for building agents. A classic example here is AutoGPT \cite{autogpt2023}. AutoGPT was the first LLM-based agent platform, and it embodies several key ideas that have informed further developments. First, it uses the idea of recursively decomposing a task into smaller tasks until these become directly executable. Second, it enables agents to access the web during problem solving---accessing web resources, and potentially actually taking actions on the web (e.g., buying a product or service).
AutoGPT attracted a huge amount of attention upon its release in 2023. What is perhaps most interesting about it is the idea that it transforms an LLM from being a purely passive and disembodied thing (a chatbot) into something that actually does things… something that is pro-active in exactly the way that was envisaged decades ago, in the original vision of agents.

\begin{figure}
    \centering
    % \includegraphics[width=0.27\linewidth]{figures/agent-architecture-llm-brain.jpg}
    % \hspace{0.1cm}
    % \vrule
    % \hspace{0.3cm}
    \includegraphics[width=0.7\linewidth]{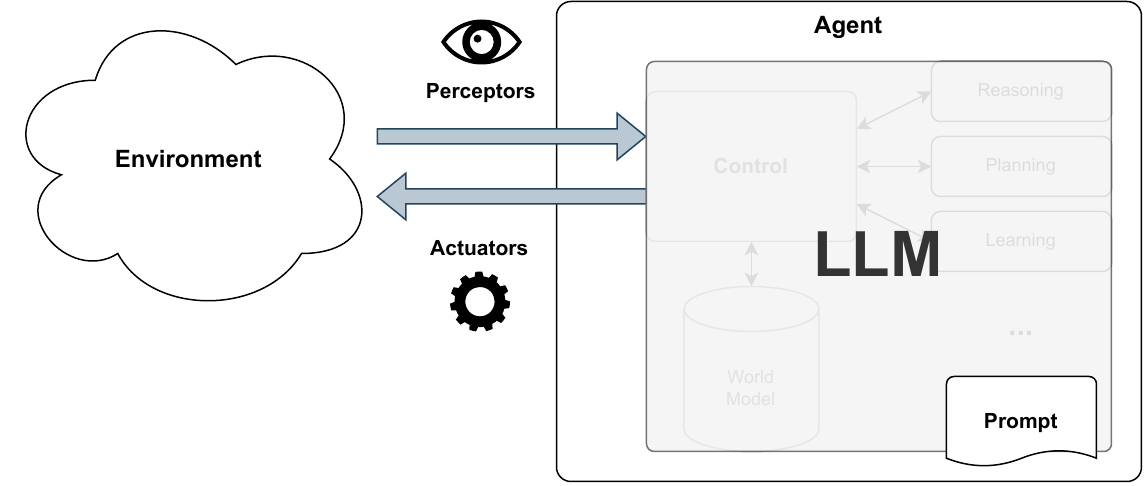}
    \caption{Can LLMs provide the ``brain'' for agents?}
    \label{fig-llmagent}
\end{figure}

\subsection{Applications of LLM-based Agents}\label{current-state-of-the-art}

\paragraph{LLM-based multi-agent systems.}
In this area, researchers aim to
use LLMs as agents \cite{Chen2023} to solve challenging problems,
leveraging different communication systems \cite{Taicheng2024}, \cite{Yongchao2023}. The idea that LLMs can be deployed in multi-agent systems led to the
development of frameworks to model their behaviours and interactions
\cite{Zhou2023}, \cite{Durant2024}. Application areas span code
generation \cite{Holt2023}, reasoning \cite{Wang2023}, \cite{Li2023} and synthetic data generation. Researchers have already
started exploring both collaborative and adversarial settings \cite{Talebirad2023}, \cite{Liang2023}, with applications to planning
\cite{Qiao2024}, reasoning \cite{Sun2023}, and debating \cite{Du2023}.
Recent work has explored the emergence of efficient, unsupervised
communication systems between agents \cite{Wu2023}, without
constraining their interactions by pre-determined linguistic constructs.

\paragraph{Multi-agent problem solving.}
This area covers general problems
that LLMs, implemented as multi-agent systems can solve. The underlying intuition is
that single or multiple LLMs can impersonate agents with different
characteristics \cite{Deshpande2023} and demographics \cite{Garimella2022} to solve problems that require domain
knowledge and adaptation. In this spirit, recent work has studied code
generation and code assistants \cite{Hong2023}, as well as data
analysis \cite{Rasheed2024}. LLMs have also been applied in
multi-agent collaboration scenarios, to shed light on the problem of
\emph{embodiment} \cite{Zhang2023}, \cite{Dasgupta2023} as well as
a tool to foster automatic scientific discovery \cite{Boiko2023}, \cite{Ni2023}.

\paragraph{Multi-agent systems for world simulation.}
Finally, an area that has recently attracted interest is that of LLM-based multi-agent systems to simulate complex systems, with applications to social sciences \cite{Ziems2023}. There has been some interest in game-theoretic evaluations of LLMs. LLMs have been studied as rational players \cite{Fan2023}, e.g., in consensus games \cite{Chen2023b} and repeated games \cite{Akata2023}. Systematic analyses have been conducted on LLMs as (rational) players in strategic games like Werewolf \cite{Xu2023} and Diplomacy \cite{Mukobi2023}. Another relevant area of application of LLMs in multi-agent systems is Theory of Mind: LLMs can be used to study the emergence of collaborative behaviours and high-order beliefs \cite{Li2023}. Finally, LLMs have also been used to validate economic theories \cite{Li2023c, Li2023b, Horton2023}.

\section{Limitations of Recent Agent Frameworks}\label{limitations-of-recent-agent-frameworks}

Since the explosion of interest in GPT-3 there have been two major development strands that have supported the development of LLMs to support agent-based systems. The first strand has been focussed on increasing the reasoning capabilities of LLMs themselves while the second has been directed at leveraging LLM's responses to enable them to provide the internal logic for agents. The reasoning models are based on ``chain-of-thought'' (CoT) concept, which involves the LLM being trained to use extended responses that involve breaking a complex question down into a series of simpler internal reasoning steps \cite{wei2023cot}. The LLM then evaluates these steps and assimilates the results into a final answer. LLMs such as the ASI:One models \cite{fetch2025asi1} combine this approach with reinforcement learning, which is used to optimize the LLM's reasoning process to earn a reward for finding the correct answer to math or logic puzzles \cite{deepseek2025r1}. The major closed and open-source developers of LLMs have all developed these types of reasoning models for complex queries.

The second, complementary approach that has primarily been driven by the open source community has involved frameworks that enable LLMs to incorporate information from the outside world or that enable them to provide the decision-making capabilities of agents that can act autonomously. This has included the use of ``function-calling'' or ``tool-use'' that enables LLMs to access external APIs. By using the correct tool, an LLM can retrieve information from a database, perform calculations, fetch information from the internet or even act in the physical world by controlling devices such as robots or drones. This type of functionality has been implemented in many LLMs including Fetch.ai's ASI:One \cite{openai2024gpt4, anthropic2024claude, fetch2025asi1}.

There are many open-source frameworks that support the integration of LLMs with external tools. It is important to acknowledge the rapid pace of development in this space; frameworks are constantly evolving. However, some fundamental and recurrent limitations can be identified. Let us discuss these limitations and their implications:

\begin{itemize}

\item \textbf{Focus on single-agent reasoning.} The LLM community has predominantly focused on enhancing individual agent capabilities—improving reasoning, planning, and tool use within isolated agents \cite{cemri2025multiagentllmsystemsfail}. This approach fundamentally misses the core insight of classical multi-agent systems research: that the complexity and power of multi-agent systems emerge from the \emph{interactions} between agents, not merely from the sophistication of individual agents \cite{Wooldridge2009}. Current frameworks often treat multi-agent scenarios as simply multiple instances of single agents operating in parallel, rather than as fundamentally different systems requiring specialized coordination mechanisms.

\item \textbf{Neglect of communication protocols and semantics.} Classical multi-agent systems research invested heavily in developing robust communication protocols, formal ontologies, and shared meaning frameworks \cite{FIPA, OWL}. In contrast, most LLM-based frameworks rely on ad-hoc natural language communication or simple API calls between agents. While natural language offers flexibility, it lacks the precision and reliability required for complex multi-agent coordination \cite{yang2025surveyaiagentprotocols}. The absence of formal communication protocols leads to ambiguity, misunderstandings, and coordination failures that classical multi-agent systems research had already addressed through standardized agent communication languages and ontological frameworks.

\item \textbf{Insufficient attention to coordination mechanisms.} The classical multi-agent systems literature extensively studied coordination, negotiation, and cooperation mechanisms \cite{RosenscheinZlotkin1994, Fatima2014}. Current LLM frameworks largely ignore these social aspects of agency, instead relying on simple orchestration patterns or basic voting mechanisms \cite{Du2023}. This oversight is particularly problematic given that real-world applications often require sophisticated coordination strategies to manage resource conflicts, task dependencies, and competing objectives.

\item \textbf{Centralization bias.} Most existing LLM-based agent frameworks exhibit a strong bias toward centralized architectures \cite{hadfield2025multiagent,langgraph2025multiagent}, where a single orchestrator or ``meta-agent'' coordinates all other agents. While this approach simplifies development and debugging, it introduces several critical limitations that classical multi-agent systems research has long recognized:

\begin{itemize}
\item \textbf{Single point of failure}: Centralized systems are vulnerable to the failure of the central coordinator, which can bring down the entire system.
\item \textbf{Scalability bottlenecks}: As the number of agents increases, the central coordinator becomes overwhelmed, limiting system scalability.
\item \textbf{Limited autonomy}: Agents in centralized systems have reduced autonomy, contradicting the fundamental principle of agent-based systems that agents should be autonomous decision-makers.
\item \textbf{Communication overhead}: All inter-agent communication must flow through the central coordinator, creating communication bottlenecks and increasing latency.
\end{itemize}

\item \textbf{Interoperability and standardization issues.} Unlike classical multi-agent systems frameworks that emphasized standardization and interoperability through protocols like FIPA \cite{FIPA}, current LLM-based frameworks often operate in silos. Agents developed in different frameworks cannot easily communicate or collaborate, limiting the potential for creating large-scale, heterogeneous multi-agent ecosystems \cite{yang2025surveyaiagentprotocols}.

\item \textbf{Limited agent discovery and matchmaking.} Classical multi-agent systems research developed sophisticated mechanisms for agent discovery, capability matching, and dynamic coalition formation \cite{Wooldridge2009}. Current frameworks typically require manual configuration of agent interactions, limiting their ability to adapt to changing requirements or discover new collaboration opportunities.

\item \textbf{Vulnerability to adversarial attacks.} A significant downside to endowing agents with significant financial or physical resources is that they have complex behavior, that leads to many potential vulnerabilities and therefore exposes large attack surfaces \cite{cemri2025multiagentllmsystemsfail}. This vulnerability is especially problematic when agents are entrusted with sensitive information, given authority over financial transactions, or allowed to control physical devices. One early example exploit of Anthropic's Model Context Protocol (MCP) demonstrates how tools with human-readable descriptions can be subject to ``tool-poisoning'' attacks, where an attacker modifies a tool's purpose and description to surreptitiously instruct the agent to use other tools to extract sensitive data \cite{modelcontextprotocol2024}.

\item \textbf{Lack of reputation and trust mechanisms.} Classical research developed sophisticated reputation systems and trust mechanisms to enable agents to assess the reliability and trustworthiness of potential collaborators \cite{Wooldridge2009}. Current LLM frameworks largely ignore these mechanisms, making it difficult to deploy agents in open, untrusted environments where malicious agents might be present \cite{yu2025surveytrustworthyllmagents}.

\item \textbf{Absence of economic coordination mechanisms.} Classical multi-agent systems research extensively studied market mechanisms, auction protocols, and resource allocation strategies for multi-agent environments \cite{RosenscheinZlotkin1994}. Current LLM frameworks lack these economic coordination mechanisms, making it difficult to manage resource allocation, cost sharing, and incentive alignment in multi-agent scenarios.

\item \textbf{Implications for scalability and resource optimization.} The computational intensity of LLMs creates unique resource management challenges that current frameworks are poorly equipped to handle. Unlike classical multi-agent systems that could optimize resource usage through intelligent task distribution and load balancing, current LLM frameworks often result in redundant computations and inefficient resource utilization.

\end{itemize}

The limitations identified in this analysis highlight the urgent need for agent frameworks that bridge the gap between classical multi-agent systems research and modern LLM capabilities. The next section will demonstrate how the Fetch.ai agent stack addresses these fundamental limitations through a principled approach that combines the best insights from both classical multi-agent systems research and contemporary AI advances.

\section{An Architecture for Modern Multi-Agent Systems}\label{the-fetch.ai-agent-stack-solution}

The recent spread of agent-based systems, driven by the capabilities of LLMs, has produced a fragmented landscape that often neglects decades of foundational research in multi-agent systems. Current frameworks frequently exhibit a bias towards centralized architectures, neglect robust communication protocols, and lack the necessary mechanisms for trust, discovery, and economic coordination in open environments. This section presents the Fetch.ai technology stack as a cohesive, multi-layered architecture designed to address these limitations. It provides an industrial-strength foundation for the development, deployment, and operation of modern multi-agent systems, realizing the original vision of autonomous agents engaging in complex interactions and transactions with one another.

The Fetch.ai architecture is composed of: a foundational layer of on-chain services, an agentic framework for development, a platform for deployment and management, and an orchestration layer for intelligent task composition. Together, these components provide the necessary infrastructure to create autonomous agents that can be securely discovered, communicate effectively, and transact with one another in a decentralized marketplace.

\begin{figure}[t]
    \centering
    \includegraphics[width=1\textwidth]{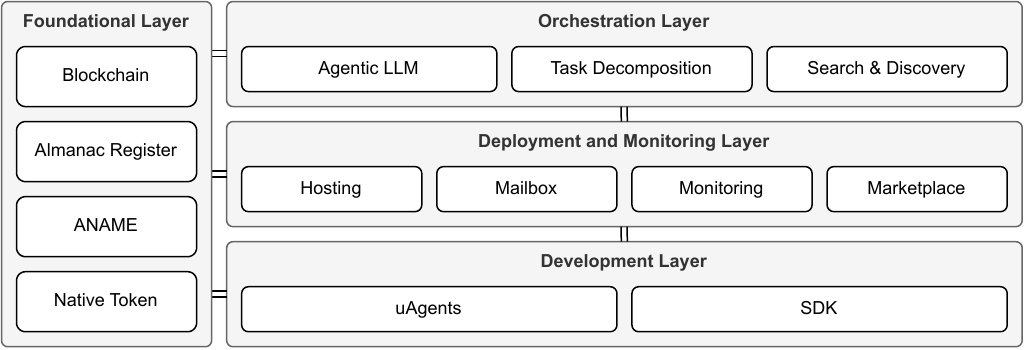}
    \caption{Architecture layers and components.}
    \label{fig:foundational_layer}
\end{figure}

\subsection{A Foundational Layer: On-Chain Trust, Discovery, and Economy}
\label{sec:foundational_layer}

To overcome the limitations of centralized and ad-hoc agent frameworks, the Fetch.ai stack is built upon a foundational layer of decentralized, on-chain services.
This layer directly addresses the critical challenges of identity, trust, discovery, and economic coordination that were identified in Section \ref{limitations-of-recent-agent-frameworks} as significant shortcomings in recent LLM-based agent systems.
By leveraging blockchain technology, this layer establishes the foundation for a robust, scalable, and censorship-resistant multi-agent ecosystem.
The on-chain nature of these core services creates significant network-backed benefits, where the utility and value of the ecosystem grow in proportion to agent participation.
The key components of this layer are the Almanac Register, the Agent Name Service (ANAME), and a native economic protocol facilitated by its native token (FET).

The \textbf{Almanac} is a smart contract deployed on the blockchain that functions as the network's \textbf{definitive and decentralized agent registry}.
It allows any agent to register its existence, services, and communication endpoint, making it discoverable without reliance on any central intermediary.
This on-chain directory is fundamental to fostering an open and permissionless environment, directly counteracting the siloed architectures and single points of failure prevalent in other frameworks.
To ensure the integrity and currency of the registry---a solution to the ``liveness problem'' in dynamic systems---the contract enforces two key mechanisms.
First, registrations are time-limited by block height, requiring agents to periodically re-register to signal their continued availability.
Second, each registration or re-registration requires the agent to prove ownership of its address by signing a unique, incrementing sequence number with its private key and submitting this signature for on-chain verification.
This proof prevents address spoofing, ensures accountability, and guarantees that the information in the registry is both current and authoritatively linked to the correct agent.

To facilitate more effective agent discovery and build greater trust between the established Web2 infrastructure and the emerging decentralized agent ecosystem, the \textbf{Agent Name Service (ANAME)} provides a mechanism for verifiably linking agents to Web2 domain names. This functionality brings a critical layer of trust from the established Web2 world, allowing agents to be associated with reputable, long-standing domains. The association process is cryptographically secure and works as follows: first, an agent's owner claims a domain they control and wishes to associate with their agent. The ANAME smart contract then issues a unique challenge, and the owner places this challenge in the domain's TXT record. An implemented set of oracle agents then verifies that the challenge has been correctly placed in the TXT record and, upon successful verification, signals the smart contract. This action creates an immutable, on-chain link, recording the association between the Web2 domain and the agent's address. This process provides a powerful reputation signal, allowing users and other agents to interact with greater confidence.

Along these services is a \textbf{native economic infrastructure}, enabled by the FET token, which facilitates a functioning on-chain economy for agents.
This protocol was specifically designed to enable micro-transactions between agents, a requirement for which traditional payment systems are ill-suited.
Each agent possesses its own wallet, allowing it to hold and transact FET autonomously.
The token is used as the primary medium of exchange for essential network services, such as paying the mandatory registration fee on the Almanac.
This fee-based structure creates a direct economic cost for consuming network resources, establishing a sustainable incentive system that discourages spam and ensures registered agents have a tangible stake in the ecosystem. This economic stake can work as a primary defense against Sybil attacks; by imposing a non-trivial cost for each registration, the system can make it prohibitively expensive for a malicious actor to create the large number of fake identities required to manipulate the network or its reputation systems.
While a testnet with a token faucet exists for development purposes, the ability to operate on the mainnet with real FET enables true commercial deployment and autonomous economic activity.
Together, these on-chain services form a cohesive and robust foundation.
The public and immutable nature of the Almanac, the user-friendly abstraction and verification of ANAME, and the economic incentives of the FET protocol create powerful network effects.
This architecture directly provides the trust, discovery, and economic coordination mechanisms that Section \ref{limitations-of-recent-agent-frameworks} identified as absent from many contemporary agent frameworks.
As more agents register and offer services, the value of the Almanac as a discovery tool increases for all participants.
This, in turn, drives further adoption and economic activity, building a self-reinforcing cycle of growth and utility that is secured by the verifiable and transparent properties of the underlying blockchain.

\subsection{The Development Layer: Core Components for Agents}
\label{the-uagent-framework}

An agentic framework suitable for industrial-scale deployment must provide developers with the core components to build agents that are not only autonomous but also secure and capable of meaningful interaction. Fetch.ai provides Python-based libraries that serve as this foundational toolkit. They are designed to address the historical challenges of interoperability and security in multi-agent systems by providing standardized, event-driven, and secure primitives for agent development. This framework enables developers to define an agent's internal logic, its communication protocols, and its security posture in a structured manner.

\subsubsection{Event-Driven and Asynchronous Architecture}

We introduce the \texttt{uAgent} framework, which is architected around an event-driven and asynchronous model, a paradigm necessary for building responsive and robust agents capable of operating in complex, networked environments. Rather than relying on a linear, blocking execution model, an agent's logic is structured as a set of functions that are triggered by specific events. These are the main components that allow for agents to percept and actuate. This is implemented through a series of Python decorators that bind functions to these events, such as:
\begin{itemize}
    \item \texttt{@on\_interval}: schedules a function to be executed periodically. This enables proactive, autonomous behavior, allowing an agent to take initiative (e.g., polling a data source) rather than only reacting to external stimuli.
    \item \texttt{@on\_message}: triggers upon the receipt of an incoming message that matches a specific data model. It represents the primary mechanism for asynchronous message handling between agents.
    \item \texttt{@on\_query}: facilitates synchronous-style request-response interactions. It processes incoming queries that conform to a specific data model and returns a response to the requester, which blocks until the response is received.
    \item \texttt{@on\_event}: captures key lifecycle events within the agent, primarily \texttt{"startup"} and \texttt{"shutdown"}, allowing for initialization and clean-up procedures.
\end{itemize}

This event-driven approach is coupled with an inherently asynchronous design, which permits an agent to manage multiple concurrent operations. The framework leverages an asynchronous event loop to handle all operations, from internal tasks to network communication. Consequently, an agent waiting for a slow network response is not prevented from performing other tasks in parallel. This non-blocking behavior is fundamental for deploying agents in distributed systems where network latency is variable.

\subsubsection{Standardized and Stateful Communication}

For heterogeneous agents to interact effectively, a common communication framework is required. The \texttt{uAgent} framework supports a full spectrum of communication needs, ranging from flexible, natural language interactions suitable for human-agent communication to formally defined protocols for unambiguous, machine-readable semantic interoperability between autonomous agents. This addresses the classical multi-agent systems challenge of achieving a shared understanding.

Let us exemplify. For a simple human-agent interaction or for flexible agent-to-agent communication, the framework provides a built-in \texttt{ChatProtocol}. This component is designed to handle natural language text, enabling developers to easily create conversational interfaces for their agents. In line with the ``preference elicitation bottleneck'' identified in previous sections, this allows users to communicate complex or nuanced goals to an agent using natural language, a capability greatly enhanced when the agent's logic is powered by an LLM.

On the other hand, for stricter agent-to-agent communication, the framework relies on a more structured approach. At the foundational level, the structure of any message is defined using a \texttt{Model} class, which inherits from Pydantic's \texttt{BaseModel} for data validation. This ensures that both sender and receiver have a consistent, machine-readable schema for the data being exchanged. This approach does not rely on an LLM's implicit understanding of terms; rather, it enforces a shared meaning through a formal, verifiable structure. Although an agent's internal context and decision-making may be powered by an LLM, as seen in recent trends, its external communication with other agents is governed by these stricter protocols.

For more complex, stateful, or standardized interactions, the framework's \texttt{Protocols} allow for any reusable collection that encapsulates a set of related message types (\texttt{Model}) and their corresponding handlers. It functions as a behavioral template, defining a specific capability or conversational pattern. For example, a developer could define a ``booking'' protocol that includes a \texttt{BookTableRequest} model, a \texttt{BookTableResponse} model, and the handler logic for the interaction. This handler can maintain state using the \texttt{uAgent} built-in context storage functionality, for instance, to record which tables have been booked.

An agent can then include this protocol in its configuration, signaling that it supports the defined interaction. When an agent broadcasts a request associated with a protocol's unique digest, it can discover and interact with all other agents on the network that have registered themselves with the same protocol digest in the Almanac registry. This mechanism provides a scalable and decentralized solution for service discovery and semantic interoperability, where agents discover each other based on a shared commitment to a formal communication contract.

\begin{figure}[t]
    \centering
    \includegraphics[width=0.8\textwidth]{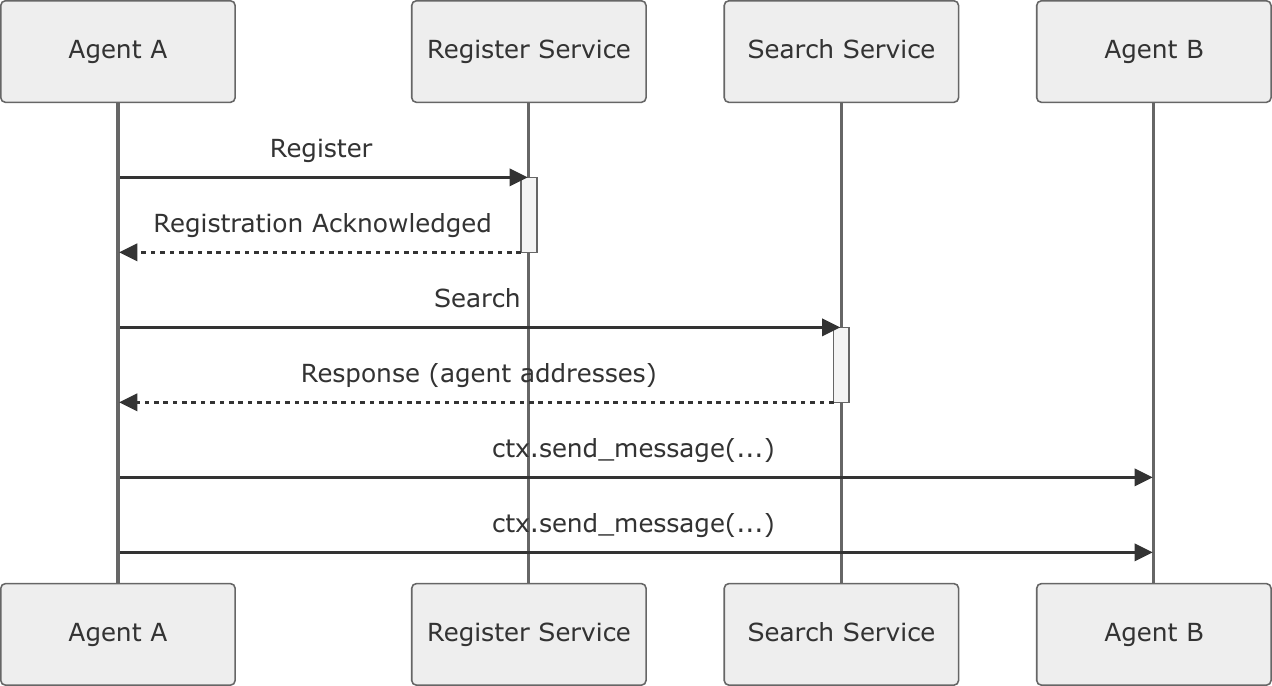}
    \caption{Agent Registration, Search, and Message Exchange.}
    \label{fig:agent_interaction_flow}
\end{figure}

\paragraph{Verifiable Identity and Messaging.}
Autonomous interaction, particularly where value or sensitive data is exchanged, is predicated on robust security mechanisms. The \texttt{uAgent} framework integrates cryptographic security at its core by providing every agent with a unique, verifiable identity and ensuring the integrity of all communications. Each agent is initialized with a unique \texttt{seed} phrase, which deterministically generates its private key that controls both the agent's address and its wallet. The agent's address and wallet are public keys. This identity enables the agent to cryptographically sign its messages, providing non-repudiable proof of origin.
A successful verification provides the recipient with a cryptographic guarantee that the message originated from the claimed sender and that its content has not been altered in transit. This integrated mechanism for message authenticity and integrity is a prerequisite for establishing trust in open and decentralized agent ecosystems.

\subsection{A Deployment, Hosting, and Monitoring Layer: The Agentverse Platform}\label{the-agentverse-cloud-agent-deployment-network}

While the \texttt{uAgent} framework provides the foundational software components for agent creation, deploying, managing, and ensuring the continuous operation of these agents presents a distinct set of infrastructural challenges. To address this, the Fetch.ai stack includes the \textbf{Agentverse}, a cloud-based platform designed to abstract away the complexities of hosting and infrastructure. It is important to note that the Agentverse is a convenience-oriented layer built on top of the decentralized foundational services; developers can always interact directly with the on-chain Almanac for discovery and registration.

The Agentverse provides a scalable and accessible environment for running agents, featuring an integrated development environment (IDE) that allows developers to write, deploy, and manage agent code from a single interface. From the hosting perspective, a primary function is providing a managed cloud, ensuring continuous uptime without requiring direct server management. Each agent operates within an isolated environment, enhancing security. for both the individual agent and the broader network. This isolation mitigates risks such as the unintended leakage of sensitive information or API keys. The platform ensures continuous uptime, meaning that once an agent is deployed, it remains online and responsive without requiring manual restarts or direct server management from the developer. This managed approach, which also supports framework-agnostic development, removes the need for developers to provision and maintain their own infrastructure, significantly lowering the barrier to entry for deploying robust, enterprise-grade agents.

Complimentary to the deployment and hosting capabilities, the platform also provides integrated tools for observability and monitoring, including logging and analytics, which are key for debugging and performance tuning. The platform provides developers with an in-browser logging console and analytics capabilities to monitor the activity and performance of their agents. These tools offer crucial insights into message flows, resource consumption, and overall agent behavior, which are key for debugging, performance tuning, and ensuring operational reliability.

As an additional, the Agentverse functions as a user-facing marketplace that provides an enhanced interface to the on-chain agent registry. While the Almanac smart contract remains the single source of truth, the Agentverse offers specialized search functionality on a centralized, cached copy of the registry data, balancing the immutability of the decentralized ledger with a more performant and rich user experience for discovery, with an advanced search engine that allows users to filter agents based on capabilities, location, and other metadata.

Furthermore, the Agentverse architecture is designed to ensure communication resilience. To solve for agents that may not maintain a persistent online presence (e.g., those on edge devices), the Agentverse offers a \texttt{Mailbox} service. This functions as a dedicated, persistent message queue for an agent. When an agent is offline, incoming messages are securely stored in its mailbox. Upon reconnecting, the agent can retrieve this backlog, ensuring no communications are lost due to intermittent connectivity. This store-and-forward mechanism is fundamental for enabling reliable asynchronous communication in real-world multi-agent systems.

\subsection{The Orchestration Layer: Intelligent Task Composition}
\label{the-orchestration-layer}

The preceding layers provide the foundational and operational components for a decentralized multi-agent ecosystem: on-chain services for trust, a secure framework for agent development, and a managed platform for deployment. The final layer of the Fetch.ai stack is the orchestration layer, which introduces an agentic LLM as the intelligent coordinator that synthesizes the capabilities of the underlying infrastructure to solve complex, high-level tasks. Like the Agentverse, this layer represents powerful product built upon the open, decentralized network. This layer can be seen as another effort to address the ``preference elicitation bottleneck'' by translating nuanced, natural language user goals into verifiable, executable, multi-agent workflows, serving as the primary cognitive interface to the network's collective capabilities.

At the core of this layer is ASI:One, a proprietary, closed-source LLM by Fetch.ai, specifically designed and optimized for agentic workflows. The model and its orchestration system are accessible via an API and the web, which is offered through a tiered model that includes a free tier with limited usage and paid tiers for more extensive access. Unlike general-purpose LLMs, ASI:One is engineered with inherent capabilities for planning, reasoning, and utilizing knowledge graphs to deconstruct abstract user objectives into discrete, actionable steps. The model is trained not only to comprehend natural language but also to interact directly with the on-chain components of the foundational layer, such as the Almanac registry, making it a ``Web3-native'' model in practice. Its reasoning process, enhanced through techniques that allow for multi-step task decomposition and refined via reinforcement learning, positions it as a cognitive engine for composing and executing plans within a multi-agent system.

A key function of ASI:One is its ability to dynamically discover and orchestrate heterogeneous agents to fulfill user requests. When presented with a high-level goal, ASI:One queries the Agentverse, and consequently the underlying Almanac registry, to identify and select the most suitable agents or tools available on the network. This discovery process leverages the rich, structured metadata that agents provide upon registration, including their capabilities and supported communication protocols, while taking advantage of advanced search features implemented in by the Agentverse marketplace.

By leveraging its advanced natural language understanding capabilities, ASI:One effectively solves the preference elicitation bottleneck that constrained early multi-agent systems. In the stack's architecture, ASI:One often embodies the role of the initial point of contact for a user's client application. Users articulate complex and nuanced goals in natural language, the ASI:One interprets this intent, translating it into a concrete sequence of autonomous agent actions and service invocations. For example, a command like ``organize a team-building event for next quarter'' is decomposed by ASI:One into sub-goals for discovering and then orchestrating venue-booking, catering, and calendar-management agents. This ability to bridge the gap between high-level human intent and low-level agent execution makes the vast network of specialized services accessible and usable without requiring technical expertise from the end-user, realizing the original vision of agents as proactive, cooperative partners.

\begin{figure}[t]
    \centering
    \includegraphics[width=1\textwidth]{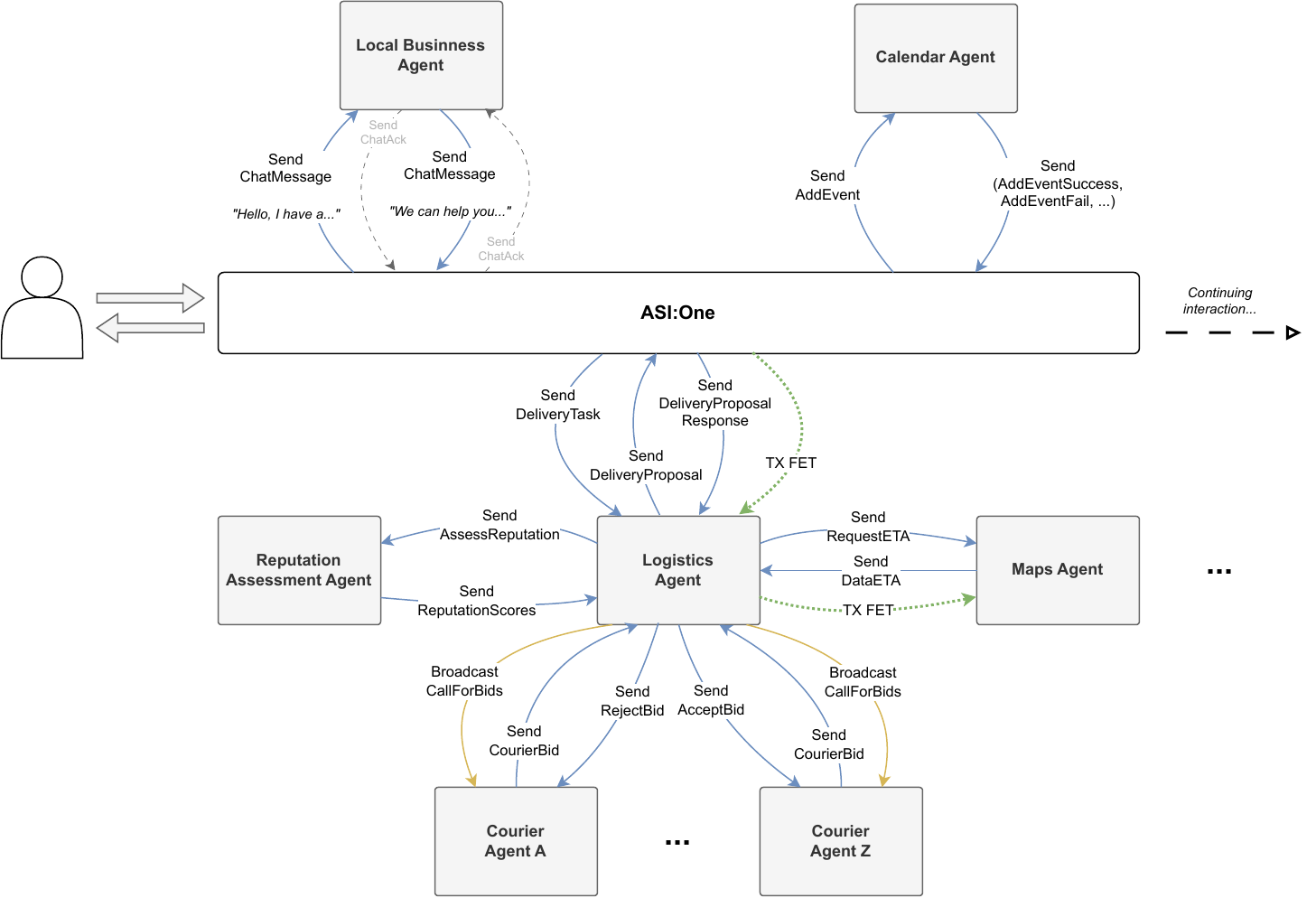}
    \caption{A multi-agent interaction flow for the logistics use case.}
    \label{fig:use_case_flow}
\end{figure}

\begin{CodeBlock}[t]
\begin{python}[frame=tb, aboveskip=1em, belowskip=1em]
ChatMessage(
    timestamp=datetime.utcnow(),
    msg_id=uuid4(),
    content=[
        TextContent(
            type="text",
            text="Hello, I have a fragile item that requires professional packaging. Can you help?"
        )
    ]
)
\end{python}
\caption{An example of a \texttt{ChatMessage} model sent by ASI:One to initiate a conversation with the Local Business Agent.}
\end{CodeBlock}

\section{Use Case: Decentralized, Autonomous Logistics}

This use case provides a practical, detailed example of the Fetch.ai architecture in action. It demonstrates how high-level user intent, expressed in natural language, is executed through a dynamic ecosystem of autonomous agents. The scenario showcases the seamless integration of a modern LLM-based agents with classic multi-agent interactions, with agents engaging in negotiation and transactions, and with discovery and communication facilitated by the Fetch.ai architecture.

In this scenario, a user interacts with a high-level application powered by Fetch.ai's agentic LLM, ASI:One, which serves as the primary interface and entry point to the agent ecosystem.
A user needs to send a valuable, fragile package from their office in Cambridge to a recipient in London. The key requirements are a balance of speed, security (careful handling), and cost-effectiveness.
Let us introduce the actors and a step-by-step breakdown of the process that enables the objective to be achieved. 

\subsection{The Actors: A Multi-Agent Ecosystem}

The scenario involves several specialized, autonomous agents, each developed using the \texttt{uAgent} framework and registered on the Fetch.ai Almanac, making them discoverable and verifiable:

\begin{itemize}
    \item \textbf{ASI:One (Orchestrator Agent):} Residing in the \textbf{Orchestration Layer}, ASI:One acts as the user's intelligent interface. It interprets the user's natural language request, decomposes the complex task into a logical sequence of sub-tasks, and orchestrates the overall workflow.
    \item \textbf{Local Business Agent (e.g., Packaging Service):} An agent representing a local business, discoverable via Agentverse. It advertises services (like packaging), location, and operating hours. As a conversational agent, it uses the \texttt{ChatProtocol} to negotiate services, answer questions, and provide quotes.
    \item \textbf{Logistics Agent:} A specialized agent responsible for managing the end-to-end delivery process. It initiates auctions for delivery tasks, evaluates bids based on multiple criteria, and ensures the contract is fulfilled.
    \item \textbf{Courier Agents:} A fleet of independent agents representing various delivery services (e.g., bike messengers, drone operators, van couriers). Each agent has unique capabilities (speed, cost, capacity, service area).
    \item \textbf{Reputation Assessment Agent:} An LLM-powered agent that assesses the trustworthiness of other agents by performing deep, qualitative analysis of public user reviews and historical performance data.
\end{itemize}

\begin{CodeBlock}[b]
\begin{python}[frame=tb, aboveskip=1em, belowskip=1em]
class DeliveryTask(Model):
    """
    Represents the initial delivery request.
    """
    source: str = Field(description="The pickup location for the package.")
    destination: str = Field(description="The delivery location for the package.")
    deadline: str = Field(description="The delivery deadline in ISO 8601 format.")
    requirements: list[str] = Field(
        description="A list of special handling requirements, e.g., ['fragile', 'keep upwards']."
    )
\end{python}
\caption{The \texttt{DeliveryTask} message model.}
\end{CodeBlock}

\subsection{The Process: From User Request to Delivery}

The following step-by-step breakdown illustrates the multi-agent interaction, with code fragments. A high-level view of this process is shown in Figure \ref{fig:use_case_flow}.

\subsubsection*{Step 1: User Intent and Task Decomposition}

The process begins with the user's natural language request to the ASI:One powered application:

\begin{displayquote}
\emph{
    ``I need to send a package from my office in Cambridge to Liverpool Street, London. It's a fragile item, so it needs careful handling. I need it delivered by 5 PM today.''
}
\end{displayquote}

\begin{figure}[t]
    \centering
    \includegraphics[width=0.95\textwidth]{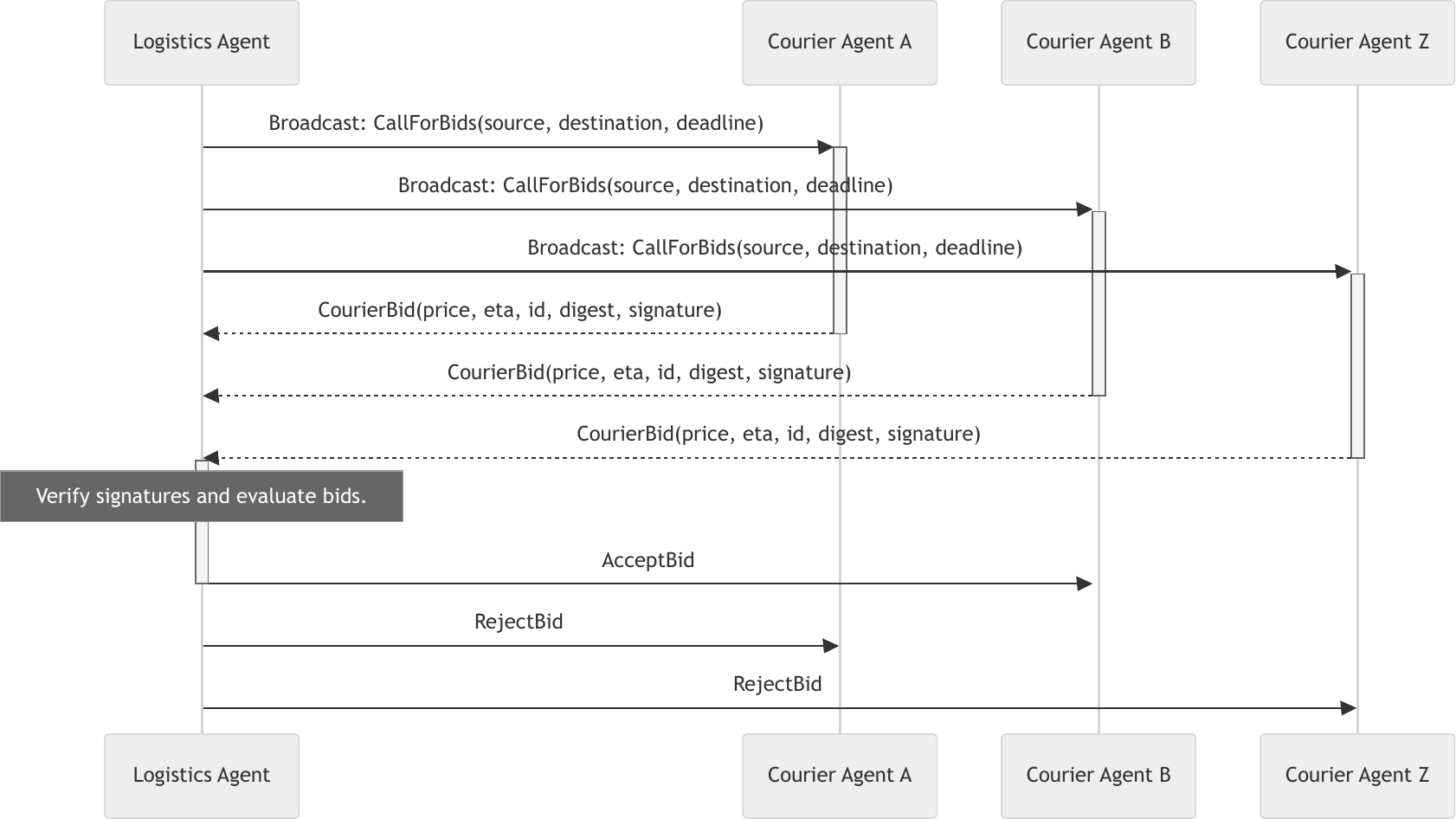}
    \caption{A one-round example of the auction protocol for courier delivery.}
    \label{fig:auction_protocol}
\end{figure}

\begin{CodeBlock}[t]
\begin{python}[frame=tb, aboveskip=1em, belowskip=1em]
# The Logistics Agent announces the task
class CallForBids(Model):
    source: str
    destination: str
    deadline: str

# Couriers respond with their signed bids
class CourierBid(Model):
    price_fet: float
    eta_minutes: int
    courier_id: str
    digest: str      # SHA-256 hash of the bid data
    signature: str   # Digital signature of the digest

# The Logistics Agent accepts the winning bid
class AcceptBid(Model): pass

# The Logistics Agent rejects all other bids
class RejectBid(Model): pass

# The protocol groups these messages
auction_protocol = Protocol("CourierAuction", version="1.1")
\end{python}
\caption{The auction protocol used by the Logistics Agent and the Couriers for bidding.}
\end{CodeBlock}

ASI:One parses this request, identifies the key parameters (e.g., source, destination, deadline, special requirements) and extracts the objective. It understands that ``fragile'' and ``careful handling'' imply a multi-step process that should begin with securing the item professionally. It also queries for agents that can manage the logistics in preparation for a later step. This search can be done via the basic Almanac register API, but for more complex needs, it can also use the Agentverse platform, which provides a more advanced search mechanism that allows for customizable search queries and filtering. This initial step showcases how ASI:One abstracts away complexity; the user simply states their need, and ASI:One independently unfolds an execution plan.

\subsubsection*{Step 2: Finding and Engaging a Packaging Service}

Recognizing the fragile nature of the item, ASI:One determines that securing professional packaging is a prerequisite. This provides a key example of Agentverse's advanced search capabilities: ASI:One performs a geolocation-based search to find business agents registered with the packaging service type near the user's office. Once a suitable Local Business Agent is found, ASI:One initiates a dialogue using the standard \texttt{ChatProtocol}.

The conversation starts with a general inquiry. The Local Business Agent would receive this, send an acknowledgement, and then could reply with a clarifying question, e.g., ``Of course. To provide a quote, could you tell me the item's dimensions and weight?''. ASI:One would then present this question to the user, forward the answer back to the Local Business Agent, and continue the conversation until a quote is finalized. Here, ASI:One acts as a seamless intermediary, managing the protocol-based dialogue with the Local Business Agent while presenting simple, actionable questions to the human user. After the conversational negotiation concludes, ASI:One relays the final quote to the user for approval.

\begin{displayquote}
\emph{
    ``After a brief chat with 'Cambridge Secure Packaging', they can professionally package your fragile item for 7 FET. Do you approve?''
}
\end{displayquote}

\subsubsection*{Step 3: Logistics Agent is Triggered}

Once the user approves the packaging cost, ASI:One proceeds with the main delivery task. It contacts the Logistics Agent it found in Step 1. The initial \texttt{DeliveryTask} message is now enriched with the information that the item will be professionally packaged, which is a critical detail for the courier auction that follows.

\subsubsection*{Step 4: Courier Agents Bid and Sign}

\begin{CodeBlock}[h]
\begin{python}[frame=tb, aboveskip=1em, belowskip=1em]
# inside courier_agent.py
import hashlib
from uagents.crypto import Identity

# ...

courier_agent = Agent(
    name="SpeedyVanCouriers", 
    seed="a_very_secret_seed_phrase"
)
courier_agent.include(auction_protocol)

@auction_protocol.on_message(model=CallForBids, replies=CourierBid)
async def handle_call_for_bids(ctx: Context, sender: str, msg: CallForBids):
    ctx.logger.info(f"Received call for bids from {sender}.")
    bid = analyze_bid(msg)  # Internal decision logic to determine bid feasibility
    if bid:
        ctx.logger.info("Task is feasible. Sending signed bid.")
        digest = encode(bid)  # Creates a digest from the bid data
        await ctx.send(sender, CourierBid(
            price_fet=bid["price"], 
            eta_minutes=bid["eta"], 
            courier_id=ctx.name,
            digest=digest.hex(),
            signature=courier_agent.sign_digest(digest)
        ))

@auction_protocol.on_message(model=AcceptBid)
async def handle_accept(ctx: Context, sender: str, msg: AcceptBid):
    ctx.logger.info(f"Our bid was accepted by {sender}! Initiating delivery.")
    # ... (Logic to initiate delivery process)

@auction_protocol.on_message(model=RejectBid)
async def handle_reject(ctx: Context, sender: str, msg: RejectBid):
    ctx.logger.info(f"Our bid was rejected by {sender}.")
    # ... (Logic to handle rejection, bid again?)
\end{python}
\caption{Courier agent logic for creating and sending a signed message.}
\end{CodeBlock}

The logistics agent receives the task and initiates a \textit{Contract Net Protocol}-style auction. To ensure shared understanding, all potential bidders implement a formal protocol (Figure \ref{fig:auction_protocol}). This negotiation is triggered and managed autonomously by the Logistics Agent, with a multi-agent auction involving agents that implement the protocol, with no human involved in the process. A crucial aspect we showcase here is message integrity and authenticity. Every bid sent by a courier is cryptographically signed using the agent's private key, and this signature is verified by the Logistics Agent. This ensures that bids are legitimate, have not been tampered with. The \texttt{CourierBid} model is extended to include a \texttt{digest} and \texttt{signature}.

The broadcasted \texttt{CallForBids} is received by all protocol-compliant couriers. Each agent autonomously evaluates the task, calculates a bid, generates a digest of the bid, and signs it.

\subsubsection*{Step 5: Bid Verification}

As signed bids arrive, the Logistics Agent's first action is to verify their authenticity. It recalculates the digest from the received bid data and uses the sender's address to verify the signature. Any bid that fails this check is immediately discarded as it could either be fraudulent or corrupted. This is a critical first-pass filter that builds a foundation of trust.

\begin{CodeBlock}[h!]
\begin{python}[frame=tb, aboveskip=1em, belowskip=1em]
# inside logistics_agent.py

@auction_protocol.on_message(model=CourierBid)
async def handle_bid(ctx: Context, sender: str, msg: CourierBid):
    ctx.logger.info(f"Received bid from {sender}: {msg.price_fet} FET")

    # Re-calculate the digest from the message payload to ensure it wasn't tampered with
    digest_from_payload = encode_bid(msg.price_fet, msg.eta_minutes, msg.courier_id)

    # Use the Identity class to verify the signature
    if not Identity.verify_digest(sender, digest_from_payload, bytes.fromhex(msg.signature)):
        ctx.logger.warning(f"Signature verification FAILED for bid from {sender}. Discarding.")
        # ... (Logic to handle invalid bids, e.g., notify sender, log error)

    ctx.logger.info(f"Signature VERIFIED for bid from {sender}.")

    # Store the now-verified bid for later evaluation
    bids = ctx.storage.get("bids", {})
    bids[sender] = { "price_fet": msg.price_fet, "eta_minutes": msg.eta_minutes, "courier_id": msg.courier_id }
    ctx.storage.set("bids", bids)

    # Subsequent logic (like triggering reputation assessment after a timeout) would follow...
\end{python}
\caption{Logistics Agent verifying an incoming bid.}
\end{CodeBlock}

\begin{CodeBlock}[t!]
\begin{python}[frame=tb, aboveskip=1em, belowskip=1em]
# inside reputation_agent.py
# ...

async def gather_public_data(address: str) -> dict: 
    # In a real implementation, this would involve API calls to search engines
    # or web scraping libraries to find publicly available information.
    return {"reviews": "...", "history": "..."}

async def llm_reputation_analysis(data: dict) -> float:
    # A real implementation would use a client library like or HTTP requests
    # to call a large language model. The prompt would instruct the LLM
    # to analyze sentiment, identify key themes in the reviews, 
    # and generate a trust score from 0.0 to 1.0.
    return {"summary": "...", "score": ...}

@reputation_agent.on_message(model=AssessReputation, replies=ReputationScores)
async def assess_reputation(ctx: Context, sender: str, msg: AssessReputation):
    ctx.logger.info(f"Received reputation request from {sender}.")
    scores = {}
    for address in msg.courier_addresses:
        public_data = await gather_public_data(address)
        score = await llm_reputation_analysis(public_data)
        scores[address] = score
    
    await ctx.send(sender, ReputationScores(scores=scores))
\end{python}
\caption{The Reputation Assessment Agent's LLM-powered analysis and decision logic.}
\end{CodeBlock}

\subsubsection*{Step 6: Trust Assessment}

For all cryptographically verified bids, the Logistics Agent proceeds to a higher-level trust assessment. This is where the power of LLMs is showcased. It triggers a specialized Reputation Assessment Agent, sending it the addresses of the valid bidders. This agent's core function is not just to collect data, but to perform a deep, qualitative analysis of it. A Reputation Assessment Agent implementation could comprise the use of tools and LLMs to:

\begin{itemize}
    \item Scrape public web data for customer reviews and historical performance records associated with each courier.
    \item Employ advanced sentiment analysis to understand the context and nuance of the reviews, distinguishing genuine feedback from noise.
    \item Synthesize this unstructured text data into a clear, actionable trust score, providing a more holistic view of a courier's reliability than simple quantitative metrics.
\end{itemize}

\subsubsection*{Step 7: Courier Selection and Proposal}

After receiving bids and trust scores, the Logistics Agent evaluates the combined data to select the best courier (e.g., balancing price, speed, and reputation). It does not finalize the deal, but instead sends a proposal to the orchestrator, ASI:One.

\begin{CodeBlock}
\begin{python}[frame=tb, aboveskip=1em, belowskip=1em]
# inside logistics_agent.py
# (verified bids are stored in ctx.storage)

@logistics_protocol.on_message(model=ReputationScores)
async def select_winner_and_propose(ctx: Context, _sender: str, msg: ReputationScores):
    best_bidder_address = ...  # abstracted utility function to select the best bidder
    winning_bid = ...

    # Store winner and losers for the final confirmation step
    ctx.storage.set("winner", best_bidder_address)
    ctx.storage.set("losers", ...)

    ctx.logger.info(f"Winner selected. Proposing to user via orchestrator.")
    await ctx.send(original_requester, DeliveryProposal(
        price_fet=winning_bid["price"],
        eta_minutes=winning_bid["eta"],
        courier_id=winning_bid["courier_id"]
    ))
\end{python}
\caption{Logistics Agent proposing the winning bid.}
\end{CodeBlock}

\subsubsection*{Step 8: Confirmation and Escrow Setup}

The Logistics Agent reports its finding to ASI:One by sending a \texttt{DeliveryProposal}.

\begin{CodeBlock}[b]
\begin{python}[frame=tb, aboveskip=1em, belowskip=1em]
# From Logistics Agent to ASI:One
class DeliveryProposal(Model):
    price_fet: float
    eta_minutes: int
    courier_id: str

# From ASI:One to Logistics Agent
class DeliveryProposalResponse(Model):
    accept: bool
\end{python}
\caption{Example of Proposal and Response models for final confirmation.}
\end{CodeBlock}

ASI:One receives the proposal and interacts with the user in natural language to get approval:

\begin{displayquote}
\emph{
  "A logistics agent has found a courier that can deliver your package by 4:30 PM for 25 FET. The courier is registered under the domain \texttt{speedyvan.example.agent}, which is verified by the ANAME service. If you approve, the funds will be held in a secure on-chain escrow contract and only released upon successful delivery. Do you want to proceed?"
}
\end{displayquote}

Upon receiving the user's approval, ASI:One instructs the Logistics Agent to finalize the agreement. The Logistics Agent then initiates the creation of an on-chain escrow smart contract.
This contract locks the 25 FET payment from the user and the terms of the delivery in a secure and tamper-proof manner. 
Only upon successful delivery confirmation will the contract automatically release the funds to the courier. Concurrently, the Logistics Agent sends \texttt{AcceptBid} to the winning courier and \texttt{RejectBid} to all others, thus closing the auction loop.
Finally, ASI:One provides a simple summary to the user:

\begin{displayquote}
\emph{
    ``Great! I've confirmed the delivery and the payment has been secured in an escrow smart contract. The SpeedyVanCouriers, a highly-rated service, will deliver your package by 4:30 PM. I will notify you upon completion.''
}
\end{displayquote}

\subsection{An Extended Scenario}

The primary goal of this example is to demonstrate a practical application of Fetch.ai's technology, showing how autonomous agents can interact with each other and with external systems to provide a seamless user experience. The scenario provides a backbone and skeleton, deliberately keeping the level of detail and code high-level for this paper's format, while still illustrating the powerful capabilities of the Fetch.ai ecosystem. The agents themselves can have their own internal logic, decision-making processes, and integrations, such as with LLMs and MCP servers. The following points outline potential extensions that could build upon this foundation:

\begin{itemize}
    \item \textbf{Pre-transaction wallet verification:} Before finalizing the agreement in Step 8, ASI:One could connect to the user's wallet to ensure they have sufficient funds to cover the delivery and packaging costs. This would involve a simple balance check. If the funds are insufficient, ASI:One could prompt the user with a message such as \textit{``Your current balance is not enough to cover the 32 FET cost. Please top up your wallet to proceed.''}

    \item \textbf{Post-delivery feedback loop:} After the delivery is confirmed complete, ASI:One could initiate a feedback mechanism. It would prompt the user to rate the service: \textit{``Your package has been delivered. How would you rate the service from 'SpeedyVanCouriers' out of 5 stars?''}. This feedback could be published in a public register to refine its trust model for future evaluations of agents in the network, which for example could be used by a Reputation Assessment Agent later on.

    \item \textbf{Recipient coordination:} To further enhance the delivery process, ASI:One could contact an agent at the destination. For instance, if the delivery is to a business, ASI:One could find the corresponding \texttt{reception.business.agent} and send a natural language message: \textit{``Please be aware that a package for [Recipient Name] is scheduled for delivery today at approximately 4:30 PM.''}

    \item \textbf{Proactive calendar integration:} Learning from user behavior, ASI:One might infer that the user typically wants important events added to their calendar. After confirming the delivery, ASI:One could proactively interact with a specialized Calendar Agent. It would send a request to this agent to create an event, such as ``Package delivery to Baker Street,'' scheduled for 4:30 PM, which would then appear in the user's calendar application.

    \item \textbf{Dynamic ETA with a Maps Agent:} To improve the accuracy of the auction, a Maps Agent could be introduced. This agent would provide real-time traffic data and ETAs. Before selecting a winner, the Logistics Agent could query this Maps Agent for a secondary, traffic-adjusted ETA for each bidder. This would involve a micro-transaction, where the Logistics Agent pays a small amount of FET to the Maps Agent for its data. This enhanced data would allow the Logistics Agent to weigh the courier bids more effectively, potentially favoring a slightly more expensive courier with a much better real-time ETA.
\end{itemize}

The scenario presented in this chapter, while not exhaustive, demonstrates that the architecture provides the tools to manage the complexities of real-world, and potentially adversarial, economic environments.

\section{Conclusion and Future Directions}\label{conclusion-and-future-directions}

The recent popular interest in intelligent agents, mostly driven by the advent of LLMs, has created a dynamic but fragmented landscape. Many of these recent agent frameworks, however, have not incorporated the extensive experience and foundational research from the field of multi-agent systems. This has led to the influx of systems with significant limitations, including a bias toward centralized architectures, a lack of formal communication protocols, and insufficient mechanisms for coordination and trust. This paper has argued that the Fetch.ai technology stack provides a cohesive, industrial-strength architecture designed to bridge this gap. By integrating insights from classical multi-agent systems research with the capabilities of modern LLMs, it offers a principled solution for the development, deployment, and operation of sophisticated multi-agent systems.

The architecture's foundational layer of on-chain services directly confronts the critical challenges of identity, discovery, trust, and economic coordination that are absent in many recent frameworks. Leveraging decentralization and cryptography provides substantial benefits for building robust agentic systems. The use of a decentralized agent registry (the Almanac), a human-readable naming service (ANAME), and a native token (FET) establishes a secure, transparent, and economically viable environment for agent interaction. This foundation is complemented by a development framework that equips agents with standardized communication protocols and verifiable identities, and platforms that abstract infrastructural complexities. The framework as a whole allows for any granularity of formalism, from a flexible chat protocol with media and text exchanges, to formal structured interactions that bring reliable executions. Finally, the orchestration layer, powered by the ASI:One Agentic LLM, provides a powerful interface for translating high-level human intent into executable multi-agent workflows. To the best of our knowledge, the Fetch.ai stack is the only deployed system with such overarching infrastructure for building trusted, secure, and economically viable agentic systems. We evaluate that these are fundamental for the deployment of modern MAS and to build a solid foundation for the future applications of AI and MAS.

Looking forward, several key areas of research and development promise to further enhance the capabilities and security of such ecosystems.

\paragraph{Advanced Cryptographic Techniques.}
The integration of advanced cryptographic primitives is set to unlock new paradigms for secure and sophisticated agent interactions. Technologies such as zero-knowledge proofs (ZKPs), multi-party computation (MPC), and confidential transactions can enable complex financial agreements and functionalities. Specific applications include the governance of shared wallets by multiple agents, the issuance of verifiable credentials for agents (e.g., proving an owner's attribute like age without revealing their identity), the creation of decentralized random beacons for fair gaming applications, the ability for LLM agents to securely sign legal agreements, and the execution of private transactions that conceal sensitive details.

\paragraph{Decentralization and Redundancy.}
A critical vulnerability in many agent frameworks is the reliance on external, centralized APIs, which introduces single points of failure. The Fetch.ai architecture is designed to mitigate this risk by enabling agents to collect information from a redundant array of sources. This design ensures that the failure of any single API does not incapacitate an agent, thereby preserving its operational capabilities and overall system resilience. Future work will continue to expand these capabilities, moving towards a truly decentralized information and service layer that mirrors the robustness of other decentralized protocols, such as Coinbase's proposed X402 payment system.

\paragraph{DAOs and Decentralized Governance.}
The long-term viability and trustworthiness of large-scale multi-agent ecosystems depend on robust governance mechanisms. The principles of Decentralized Autonomous Organizations (DAOs) are highly relevant in this context. By applying insights from game theory and established agent research on protocol design and incentives, it is possible to construct DAOs capable of governing agent marketplaces, managing dispute resolution, and overseeing the evolution of the platform itself. This approach ensures that the ecosystem remains fair, transparent, and adaptable to the evolving needs of its participants.

\paragraph{The Evolution of LLMs.}
The advancement of LLMs will continue to be a primary driver of agent capabilities. Future developments are expected to address current limitations and unlock new functionalities. As pre-training on existing text datasets nears its limits due to the exhaustive mining of the internet, focus is shifting towards more advanced reasoning models. This includes models that are directly trained to use tools, rather than invoking them as a secondary step in response to a prompt. Furthermore, research into non-recursive reasoning models, potentially based on diffusion, could allow for more extended inference-time computation. A significant frontier is the development of models that incorporate generative models of the world, analogous to human thought processes. Such advancements build on the trajectory of machine learning, which has already demonstrated superhuman performance in solving narrow problems , and may be enhanced by findings that suggest LLMs operate on a shared universal linguistic geometry, akin to Platonic forms.

\pagebreak

\bibliography{references}
\bibliographystyle{plain}

\end{document}